\documentclass[12pt,preprint]{aastex}
\def\la{\mathrel{\hbox{\rlap{\hbox{\lower4pt\hbox{$\sim$}}}\hbox{$<$}}}}
\def\sun{\hbox{$\odot$}}

\begin{document}
\title{A 3\% Determination of $H_0$ at Intermediate Redshifts}
\author{J. A. S. Lima\altaffilmark{1,a} and J. V. Cunha\altaffilmark{2,b}}
\affil{$^{a}$Departamento de Astronomia, Universidade de S\~ao
Paulo,
\\ 05508-900 S\~ao Paulo, SP, Brazil\\
$^{b}$Faculdade de F\'{i}sica, Universidade Federal
do Par\'a,\\
66075 - 110, Bel\'em, PA, Brazil}
\altaffiltext{1}{limajas@astro.iag.usp.br}
\altaffiltext{2}{jvcunha@ufpa.br}

\begin{abstract}
Recent determinations of  the Hubble constant, $H_0$,  at
extremely low and very high redshifts based on the cosmic distance
ladder (grounded with trigonometric parallaxes) and a
cosmological model (applied to Planck 2013 data) respectively, are
revealing an intriguing discrepancy (nearly 9\% or 2.4$\sigma$)
that is challenging astronomers and theoretical cosmologists. In
order to shed some light on this problem, here we discuss a
new determination of $H_0$ at intermediate redshifts ($z\sim
1$), using the following four cosmic probes: (i) measurements of
the angular diameter distances (ADD) for galaxy clusters
based on the combination of Sunyaev-Zeldovich effect and X-ray
data ($0.14 \leq z \leq 0.89$), (ii) the inferred ages of old high
redshift galaxies (OHRG) ($0.62 \leq z \leq 1.70$), (iii)
measurements of the Hubble parameter $H(z)$ ($0.1 \leq z \leq
1.8$), and (iv) the baryon acoustic oscillation (BAO) signature
($z=0.35$). In our analysis, assuming a flat $\Lambda$CDM
cosmology and considering statistical plus systematic errors 
we obtain $H_0 = 74.1^{+2.2}_{-2.2}$ km s$^{-1}$ Mpc$^{-1}$
(1$\sigma$) which is a $3\%$ determination of the Hubble constant
at intermediate redshifts. We stress that each individual test
adopted here has error bars larger than the ones appearing in the
calibration of the extragalactic distance {\it ladder}. However,
the remarkable complementarity among the four tests works
efficiently in reducing greatly the possible degeneracy on
the space parameter ($\Omega_m,h$) ultimately providing a
value of $H_0$ that is in excellent agreement with the
determination using recessional velocities and distances to
nearby objects.
\end{abstract}



%{\vskip3.0cm}

\section{Introduction}

One of the most important  observational quantities for cosmology
is the Hubble constant, $H_0$, whose value determines the present
day expansion rate of the Universe. The determination of $H_0$
remains a very active research topic since  the early days of
physical cosmology (Jackson 2007; Riess {\it et al.} 2011; Suyu {\it {et al.}}
2012; Freedman {\it {et al.}} 2012). Several groups and ongoing
missions are now focusing their efforts to a high-accuracy
calibration of  $H_0$ since it works like a key to quantify many
astronomical phenomena in a wide range of cosmic scales. It also
plays an important role for several cosmological calculations as
the physical distances to objects, the age, size, and
matter-energy content of the Universe (Freedman \& Madore 2010;
Ch\'{a}ves {\it{et al.}} 2012; Farooq and Ratra 2013; Ade {\it{et
al.}} 2013).

Currently, the more robust constraints on $H_0$ are being
obtained from local tests at low redshifts (i.e., $z \ll 1$). The
primary method is based on a cosmic {\it distance ladder} interlinking different distance indicators.
The basic measurements and strategies commonly adopt: Cepheids,
tip of the red giant branch, maser galaxies, surface brightness
fluctuations, the Tully-Fisher relation, and type Ia supernovae.
In the last decade, the Hubble Space Telescope (HST) Key
Project did a magnificent job in decreasing significantly
the errors on Hubble constant (for a review see Freedman \& Madore
2010). Recently, Riess et al. (2011) used the HST observations to
determine $H_0$ from Cepheids and Supernovae (SNe Ia). Through a
rigorous analysis of the statistical and systematic errors they
obtained $H_0 = 73.8 \pm 2.4$ km s$^{-1}$ Mpc$^{-1}$ (1$\sigma$),
corresponding to a $3.3\%$ uncertainty. Later on, using
Spitzer Space Telescope, Freedman {\it{et al.}} (2012)
re-calibrated the HST Key Project sample and found $H_0 = 74.3 \pm
2.6$ km s$^{-1}$ Mpc$^{-1}$ (1$\sigma$).

The local distance
ladder may also  be susceptible to unknown sources of systematics like the possible existence of a
``Hubble bubble'' or other
local effects that would affect measurements of the nearby expansion-velocity (Jha {\it{et
al.}} 2007; Sinclair {\it{et al.}} 2010; Marra {\it{et al.}}
2013). The possibility of an observational convergence in the near
future has increased in the last few years, however, additional
progress, say, for a determination with $2\%$ uncertainty will
demand a closer scrutiny of the cosmic distance ladder (Suyu {\it{et al.}} 2012; Freedman {\it{et
al.}} 2012).

On the other hand, cosmologists desire
accurate measurements of the $H_0$ mainly to refine the
constraints on the neutrino masses ($\sum m_{\nu}$), density
($\Omega_{\Lambda}$), and the equation of state parameter
($\omega$) of dark energy based on the cosmic microwave background
(CMB) anisotropies data (Macri {\it{et al.}} 2006; Sekiguchi
{\it{et al.}} 2010). It is also widely known that CMB data alone
can not supply strong constraints on $H_0$ (Spergel {\it{et al.}}
2007, Komatsu {\it et al.} 2009), a problem closely related with
the degeneracy of the space parameter.  Different grouping of the
parameters ($H_0$, $\Omega_M$, $\Omega_{\Lambda}$, $\omega$, etc.)
produce the same prediction of the CMB anisotropies. This
degeneracy problem can be alliviated only by using a prior on $H_0$
from independent probes (Hu
2005). In this regard, Komatsu {\it{et al.}} (2011), used
CMB, Supernovae (SNe Ia) and Baryon acoustic oscillations (BAO)
data, to derive a model-dependent value of $H_0$. It should also
be recalled that the phenomenology underlying the CMB test
operates on observations made at very high redshifts ($z
\simeq$ 1070), during the decoupling of radiation and matter. 
Such determination of $H_0$ involves a
combination of physics and phenomena at very different scales and
epochs of the cosmic evolution (low, intermediate and high
redshifts), and, more important to the present article, due to the
inclusion of SNe Ia data, the derived value of $H_0$ is also
somewhat dependent on the cosmic distance ladder. Therefore, it is not only important to derive constraints on 
$H_0$ based on many different kinds of observations, but,
also to avoid the combination of phenomena occurring at very
different epochs (like SNe Ia and CMB).

Hinshaw {\it{et al.}} (2012) using CMB experiments from WMAP-9
found $H_0 = 70.0 \pm 2.2$ km s$^{-1}$ Mpc$^{-1}$ (1$\sigma$)
whereas Ade {\it{et al.}} (2013) using Planck mission
analysis, reported $H_0 = 67.4 \pm 1.4$ km s$^{-1}$ Mpc$^{-1}$
(1$\sigma$), corresponding to a 2.1\% uncertainty. It is worth
noting that the current results from different experiments
working with very high redshifts physics are consistent with
each other; however, the CMB experiments are discrepant with the
local measurements. In particular, the PLANCK mission tension is
at $2.4\sigma$ confidence level (Ade {\it{et al.}} 2013). In
the present ``era of precision cosmology'', investigating
this tension more fully may bring new insights into improving the
determination of the Hubble constant. Naturally, if such tension
is not a mere consequence of systematics, and it is further
strengthened by the incoming data, a new cosmology beyond
$\Lambda$CDM may also prove necessary.

Recently, some cosmological tests at intermediate
redshifts ($z\sim 1$) have emerged as a promising technique to
estimate $H_0$ (Simon {\it{et al.}} 2005; Deepak \& Dev 2006;
Cunha {\it et al.} 2007; Lima {\it et al.} 2009; Stern
{\it{et al.}} 2010). The main advantage
of these methods is their independence of local calibrators
(Carlstrom {\it{et al.}} 2002; Jones {\it{et al.}} 2005). In
principle, such methods provide a crosschecking for local
direct estimates which are free from Hubble bubbles, since the
majority of galaxy clusters are well inside the Hubble flow.
All these methods are dependent on the assumptions about
the astrophysical medium properties, as well as, of the
cosmological model adopted in their analysis, and, individually, are not yet competitive
with the traditional methods based on the cosmic distance ladder
(Freedman {\it{et al.}} 2001).

In this Letter, we identify a remarkable complementarity involving
four different cosmological probes at intermediate redshifts
($z\sim 1$) whose combination provides a valuable cross-check for $H_0$. The first test is based on the angular diameter distance from galaxy clusters via
Sunyaev-Zeldovich effect (SZE) combined with measurements of the
X-ray flux (SZE/X-ray technique). Although suggested long ago, only recently it has been applied for a fairly
large number of clusters (Bonamente {\it {et al.}} 2006; Cunha
{\it{et al.}} 2007; Holanda {\it{et al.}} 2012). The
second one is the age estimates of galaxies and
quasars at intermediate redshifts. It became possible through the
new optical and infrared techniques together the advent of
large telescopes (Lima {\it et al.}
2009). The third  possibility arises from measurements of the
Hubble parameter, $H(z)$, from differential ages of galaxies (Simon {\it{et al.}} 2005; Gazta\~naga
{\it{et al.}} 2009; Stern {\it{et al.}} 2010), while the fourth is the BAO signature (Eisenstein {\it et al. 2005}). 
The cooperative interaction among these independent tests
reduce greatly the errors on $H_0$, and,
more interesting, located just on a redshift zone distinct both
from CMB anisotropies and the methods defined by the cosmic
distance ladder (Cepheid, SNe Ia, etc.). As we shall see, these four probes act in concert to predict constraints on $H_0$ that are in excellent agreement with local measurements. 

%In principle, it is
%possible to derive tighter constraints on the possible values of
%$H_0$ once the cosmology is fixed. However, in our analysis we
%marginalize over the cosmology by adding the fourth test which is
%represented by the BAO signature. The advantage of a combination
%with BAO (a scale independent of $H_0$) was suggested earlier by
%Cunha, Marassi \& Lima (2007), however, restricted only to the
%Hubble constant determination based on the SZE technique. In what
%follows, we show that the four distinct probes adopted here act in
%concert to predict constraints on the value of $H_0$ which are in
%excellent agreement with local measurements.

\begin{figure*} % Duas figuras lado a lado
         {\includegraphics[width=3.4in,height=3.0in]{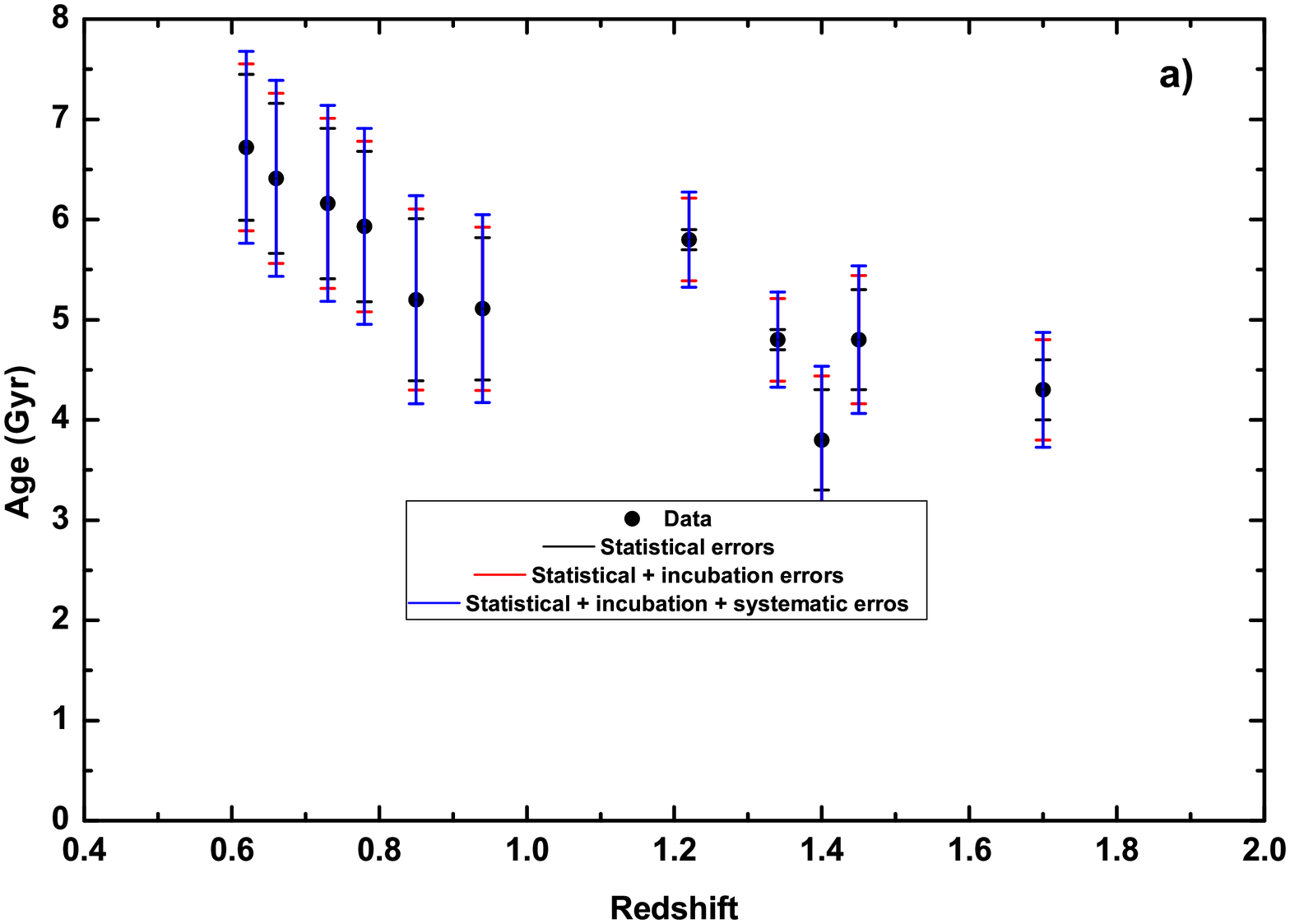}}
         {\includegraphics[width=3.4in,height=3.0in]{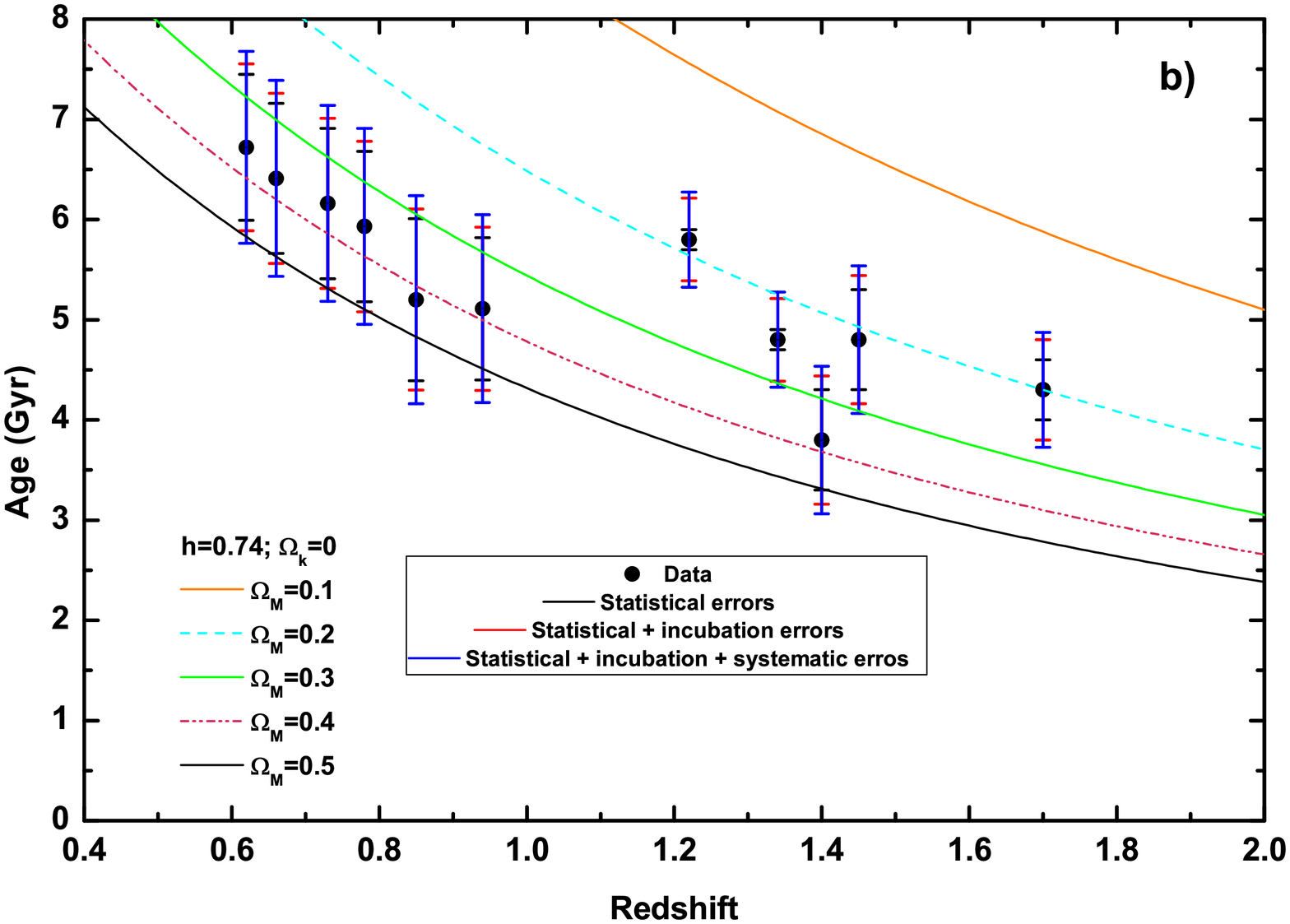}}
{\centerline{
{\includegraphics[width=3.4in,height=3.0in]{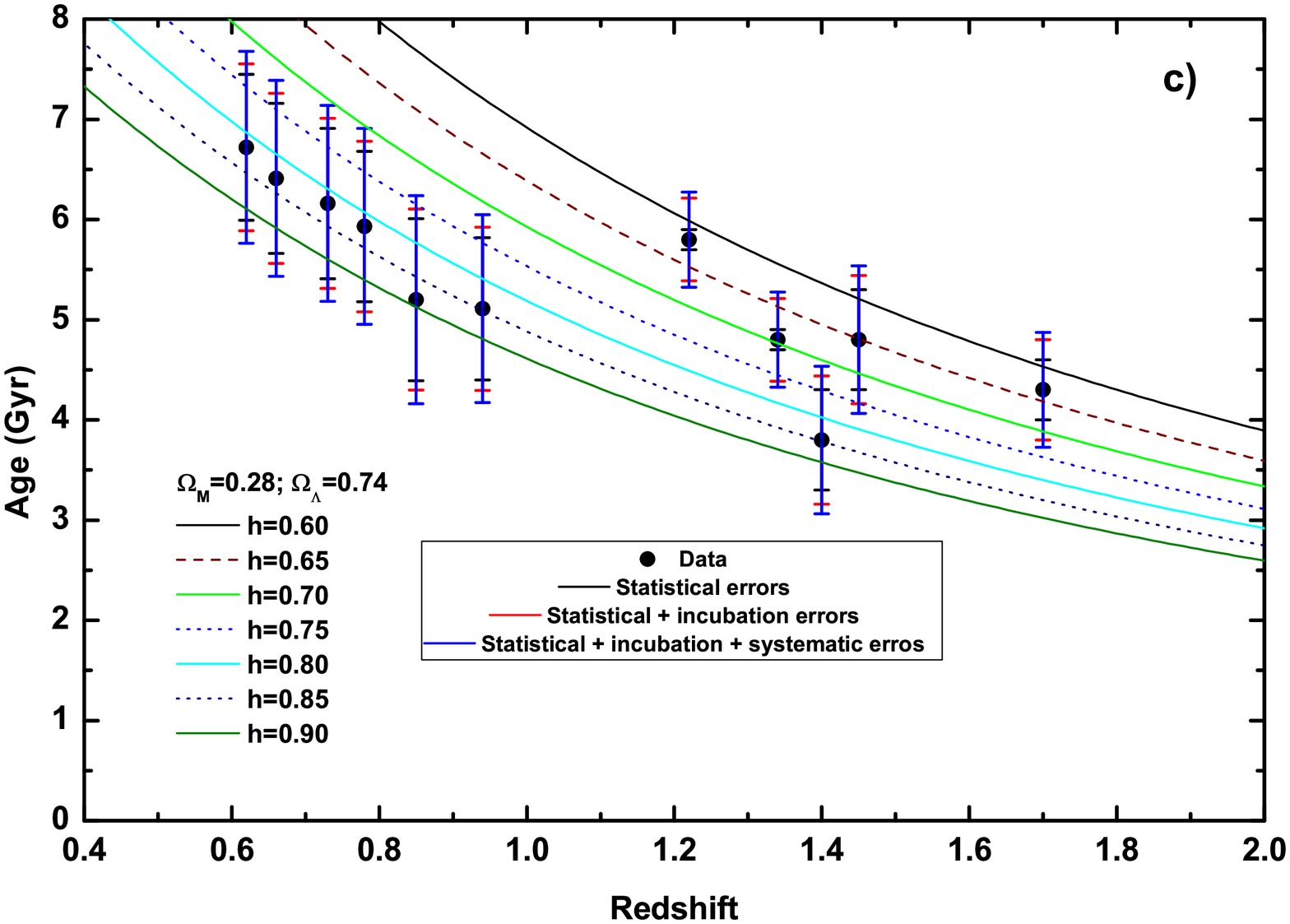}}}}
       \caption{{\small {Old galaxies data at intermediate redshifts as a cosmic probe. {\bf{a)}} Age-redshift plane and the total sample of galaxies.
       Black points for $z<1.0$ and $z>1.2$ correspond, respectively, to the Ferreras {\it{et al.}} (2009) and
       Longhetti {\it{et al.}} (2007) samples. Black, red and blue bars represent the statistical, statistical+incubation and
       statistical+incubation+systematic errors, respectively.
      Data points correspond to 11 galaxies of our selected
      subsample (see the main text). {\bf{b)}} Effect of $\Omega_M$. Age of the Universe for some selected values of the density parameter. 
      %All plots were drawn by choosing  $h$ as the best fit of our analysis ($h = 0.74$).
      For smaller values of $\Omega_M$ the ages of the Universe for a given redshift increase, thereby accommodating the oldest
      selected objects. {\bf{c)}} The $h$ effect on the Age-redshift relation. Dotted
curves are the predictions of the cosmic concordance model
($\Omega_M= 0.28, \Omega_{\Lambda}= 0.74$) and different values of
$h$. For values smaller than $h=0.6$ or bigger than
$h=0.9$ the  curves predicted by  the models move away from the
data}.}} \label{fig:X}
\end{figure*}

\section{Basic Equations, Probes and Samples}\label{sec:Samples}

%Let us now consider that the Universe is described by a flat
%Friedmann-Lema\^{\i}tre-Robertson-Walker (FLRW) geometry 
%%(in our units $c=1$)
%\begin{equation}
% ds^2=dt^2-a^2(t)\left[dr^2+r^2(d\theta^2+sin^2\theta d\phi)\right],
%\end{equation}
In the $\Lambda$CDM cosmology, the angular diameter distance (AD), ${\cal{D}}_A$),
can be written as (Lima \& Alcaniz 2002; Lima
{\it{et al.}} 2003; Holanda {\it{et al.}} 2010)
\begin{equation}
{\cal{D}}_A(z;h,\Omega_M) = \frac{3000h^{-1}}{(1 +
z)}\int_{o}^{z}\frac{dz'}{{\cal{H}}(z';\Omega_M)} \, \mbox{Mpc},
\label{eq1}
\end{equation}
where $h=H_0/100$ km s$^{-1}$ Mpc$^{-1}$ (subscript $0$ denotes
present day quantities), and the dimensionless function
${\cal{H}}(z';\Omega_M)$ is given by ${\cal{H}} = \left[\Omega_M(1
+ z')^{3} + (1 -\Omega_M)\right]^{1/2}$.

Bonamente and collaborators (2006) determined the ADD distance to
38 galaxy  clusters  in the redshift range $0.14 \leq z \leq 0.89$
using X-ray data from Chandra and Sunyaev-Zeldovich effect data
from the Owens Valley Radio Observatory and the
Berkeley-Illinois-Maryland Association interferometric arrays.
Assuming spherical symmetry, the cluster plasma and dark matter
distributions were analyzed by using a hydrostatic equilibrium
model accounting for radial variations in density, temperature and
abundances. The common statistical contributions for such galaxy
clusters sample are: $\pm 2$\% from galactic (and extragalactic) X-ray background,
galactic N$_{H}$ $\leq \pm 1\%$, CMB anisotropy $\leq \pm
2\%$, $\pm 8$\% associated to Sunyaev-Zeldovich effect
from point sources, while statistics due  to cluster
asphericity amounts to $\pm 15$\%. The estimates  of systematic effects include: X-ray temperature calibration $\pm 7.5$\%, radio
halos $+3$\%, while calibrations of SZE
and  X-ray background flux contribute, respectively, with $\pm 8$\% and $\pm 5$\%. Different authors (Mason {\it{et al.}} 2001; Reese {\it{et al.}}
2002, 2004) also believe that typical statistical errors amounts for nearly $20$\%  with + 12.4\% and - 12\%  for systematics (see table 3 in Bonamente {\it{et al.}}
2006).

On the other hand, the age-redshift relation, $t(z)$, for a flat
($\Lambda$CDM) model has also only two free parameters
($H_0,\Omega_{M}$)
\begin{eqnarray}
t(z;h,\Omega_M)& = &H_{0}^{-1}\int_{0}^{\frac{1}{(1 + z)}}{dx
\over x\sqrt{\Omega_{\rm{M}}x^{-3} + (1 - \Omega_{\rm{M}})}} .
\end{eqnarray}
Note that for $\Omega_{\rm{M}} = 1$ the above expression reduces
to the well known result for Einstein-de Sitter model (CDM,
$\Omega_{\rm{M}}=1$) for which $t(z) = \frac{2}{3}H_0^{-1}(1 +
z)^{-3/2}$. The above expression means that by fixing $t(z)$ from
observations one may derive limits on
the cosmological parameters $\Omega_{\rm{M}}$ and $H_0$ (or
equivalently $h$).  It is also worth noticing that the quantities appearing in the product defining the age parameter, $T=H_0t_z$,
are estimated based on different (independent) observations (Lima \& Alcaniz 2000;
Fria\c{c}a {\it et al.} 2005).

In this context, Ferreras {\it et al.} (2009) catalogued 228 red
galaxies on the interval $0.4 <z< 1.3$ using HST/ACS slitless
grism spectra from the PEARS program thereby studying the stellar
populations of morphologically selected early-type galaxies from
GOODS North and South fields. The first subsample adopted here
consists of six passively evolving old red galaxies selected from
the original Ferreras {\it et al.} sample. The second  data set of
old objects is also a subsample of the Longhetti {\it {et al.}}
(2007) sample. Only nine field galaxies
spectroscopically classified as early-types at $1.2 < z <1.7$ were selected from a complete sample  of the Munich Near-IR
Cluster Survey (MUNICS) with known optical  and
near-IR  photometry (the age of each galaxy estimated from its stellar population).  It should be stressed
that for both samples, the selected  data set (eleven galaxies) provide the most
accurate ages and the most restrictive galaxy ages.

In Figure 1a, we display the sample constituted by the eleven data
points chosen from  two distinct subsamples of old objects. As
discussed by Lima {\it et al.} (2009), we have added an incubation
time with a conservative error bar for all galaxies.  It is defined by the amount of time interval from the
beginning of structure formation process in the Universe until the
formation time ($t_f$) of the object itself. It will be assumed here that $t_{inc}$
varies slowly with the galaxy and redshift in our sample thereby associating a
reasonable uncertainty, $\sigma_{t_{inc}}$, in
order to account for the present ignorance on this kind of ``nuisance"
parameter (Fowler 1987; Sandage 1993). Here we
consider $t_{inc}=0.8 \pm 0.4$ Gyr, and following Lima {\it
et al.} (2009) we also combine
statistical and systematic errors. 

In Figures 1b and 1c, we compare the age of these old objects at
intermediate redshifts with the predictions  of the $\Lambda$CDM
models for different values of the free parameters ($\Omega_M$ and
$h$). It is also worth noticing that the present status of
systematic uncertainties in this context is still under debate.
Jimenez {\it {et al.}} (2004) studied sources of systematic errors
in deriving the age of a single stellar population and concluded
that they are not larger than $10-15$ per cent. Others authors
consider systematic errors around $20$\% (Percival \& Salaris
2009). In the present study we have adopted $15\%$ for all data.

The third observational probe comes from  $H(z)$ data, obtained
from differential ages of galaxies and radial BAO (Simon {\it {et
al.}} 2005; Gazta\~naga {\it {et al.}}  2009). Some years ago,
Jimenez and collaborators suggested an independent estimator for
the Hubble parameter (differential ages of galaxies) and used it
to constrain the equation of state of dark energy (Jimenez {\it
{et al.}} 2003). Later on,  the differential ages of
passively-evolving galaxies were used to obtain $H(z)$ in the
range of $0.1 < z < 1.8$ (Simon {\it {et al.}} 2005), and this
sample was further enlarged by Stern {\it {et al.}} (2010). In
addition, Gazta\~naga {\it {et al.}} (2009) took the BAO scale as
a standard ruler in the radial direction ("Peak Method") thereby
obtaining three more additional data: $H(z = 0.24) = 79.7 \pm
2.7$, $H(z = 0.34) = 83.9 \pm 3.2$, and $H(z = 0.43) = 86.5 \pm
3.5$, which are model and scale independent. Now, by comparing the
theoretical expression, $H(z)= H_0 \left[\Omega_M(1 + z)^{3} + (1
-\Omega_M)\right]^{1/2}$,  with the observational data the
corresponding  bounds can be readily derived.  We remark that the
relative age difference (the key to the method) is only of the
order of $2 - 3\%$ (Stern {\it {et al.}} 2010). However, to be
more conservative we are assuming here systematic errors of $8\%$
for the relative age difference and radial BAO data.

A joint analysis based on the above 3
different probes already provides  tight constraints on the value
of $H_0$. However, as shown by Cunha {\it{et al.}} (2007),  the
analysis of ${\cal{D}}_A$ data (from SZE/X-ray technique) leads to
more stringent constraints on the space parameter ($\Omega_M, h$)
when combined  with the BAO signature (Eisenstein {\it{et al.}}
2005; Percival {\it {et al.}} 2010), and the same happens when the
ages of high redshift objects are considered (Lima {\it et al.}
2009). Therefore,  instead to fix a definite flat $\Lambda$CDM
cosmology, it is natural to
leave $\Omega_M$ free which will be more accurately fixed by
adding the BAO signature as a fourth probe to the complete joint
analysis performed here. The remnant BAO peak can be interpreted as a consequence of the baryon acoustic
oscillations in the primordial baryon-photon plasma prior to
recombination. It was detected from a large sample of luminous red galaxies and can be characterized by a
dimensionless parameter:
\begin{eqnarray}
{\cal{A}} \equiv {\Omega_{\rm{M}}^{1/2} \over
{{\cal{H}}(z_{\rm{*}})}^{1/3}}\left[\frac{1}{z_{\rm{*}}}
\Gamma(z_*)\right]^{2/3}  = 0.469 \pm 0.017, %\nonumber
\label{A}
\end{eqnarray}
where $z_{\rm{*}} = 0.35$ is the redshift at which the acoustic
scale has been measured, and $\Gamma(z_*)$ is the dimensionless
comoving distance to $z_*$. The above quantity  is
independent of the Hubble constant, and, as such, this BAO
signature alone constrains only the $\Omega_M$ parameter.

\begin{figure*} % Duas figuras lado a lado
         {\includegraphics[width=3.4in,height=3.2in]{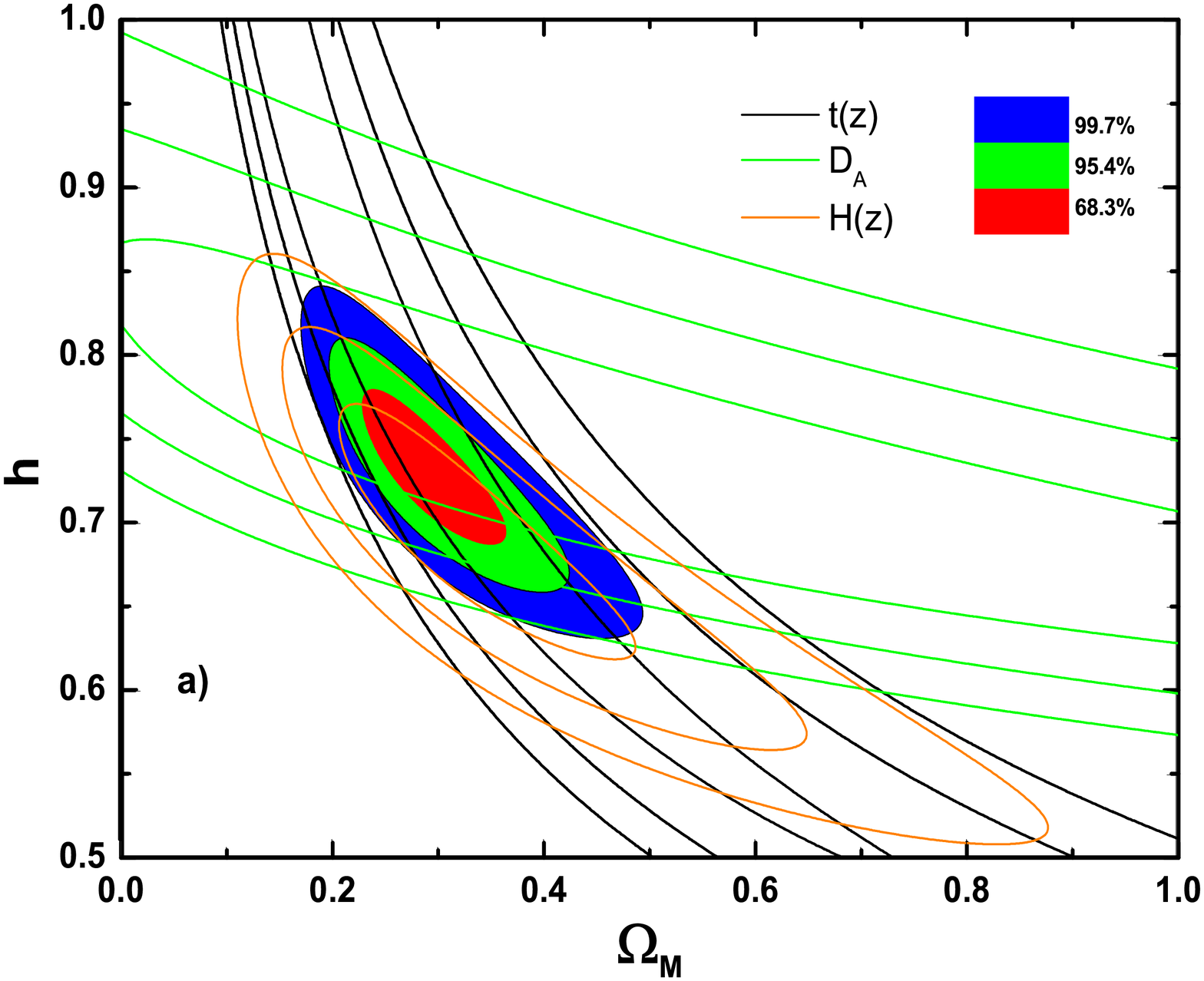}}
         {\includegraphics[width=3.4in,height=3.2in]{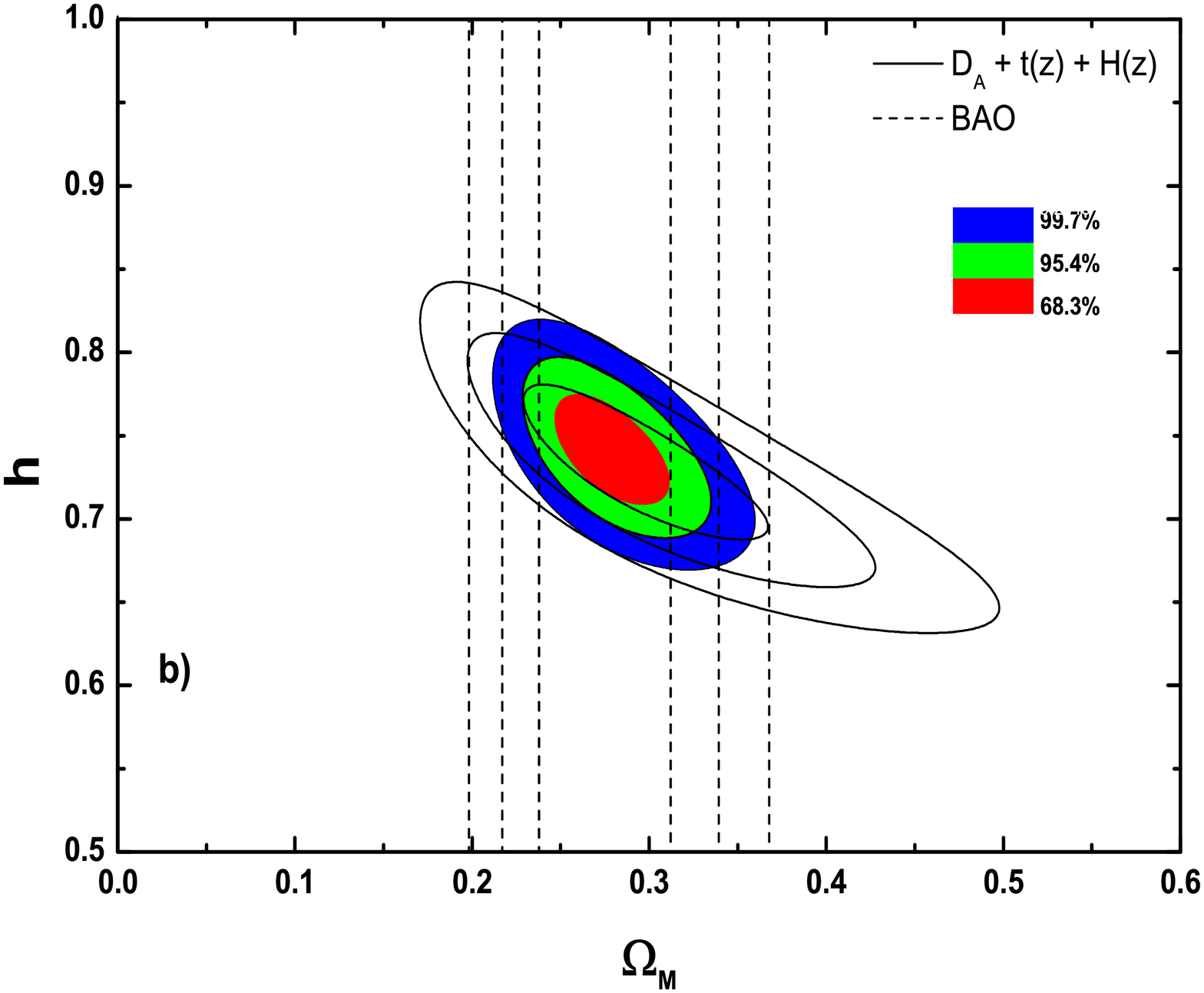}}
{\centerline{
{\includegraphics[width=3.4in,height=3.2in]{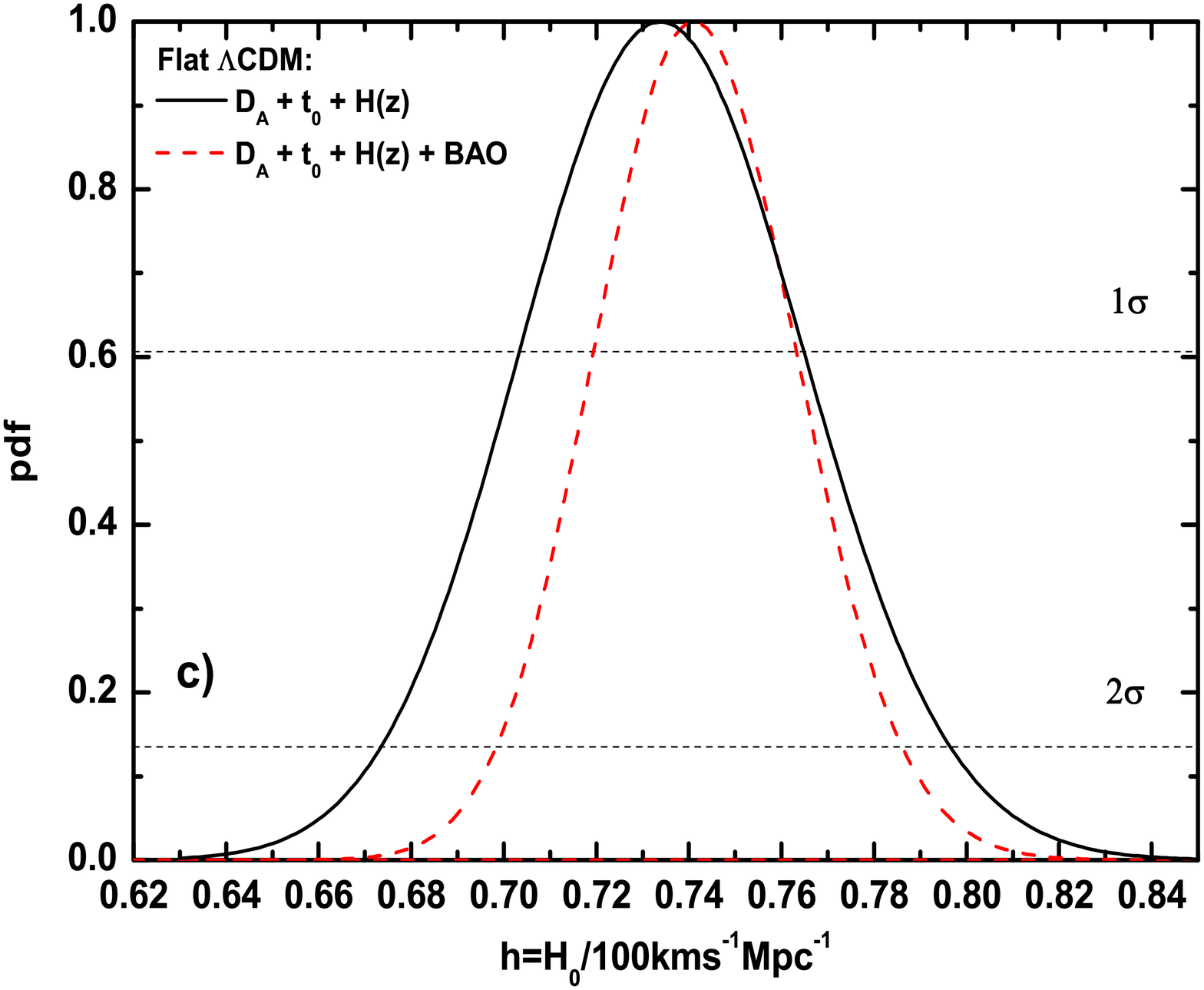}}}}
       \caption{{\small {$H_0$ determination at intermediate redshifts. \small {\bf{a)}} $H_0$ from three different probes.
       The ($\Omega_M, h$)
plane provided by the SZE/X-ray + galaxy age + $H(z)$ data
including statistical plus systematic errors. The best-fitting
values are $h=0.734$ and $\Omega_M= 0.290$. {\bf{b)}} The BAO
razor. Contours in the ($\Omega_M, h$) plane from a joint analysis
involving SZE/X-ray + galaxy age +  $H(z)$ + BAO data. The
best-fitting model converges to $h= 0.741$ and $\Omega_M= 0.278$.
{\bf{c)}} Final values of $H_0$. Likelihood functions for the $h$
parameter in a flat $\Lambda$CDM Universe. The solid black curve
corresponds to a joint analyses involving SZE/X-ray + galaxy age +
Hubble parameter whereas the red dashed curve includes the  BAO
signature. The horizontal lines are cuts in the regions of
$68.3\%$ and $95.4\%$ probability. The joint analysis performed
with the four different probes (including statistical plus
systematic errors) provides $h= 0.741\pm 0.022$ (1$\sigma$). This constraint on $H_0$ is in excellent
agreement with the latest determinations based on the
extragalactic distance ladder (Table I).}}} \label{fig:X}
\end{figure*}

\section{Complementarity for $H_0$}\label{sec:Complementarity}

Let us now consider a joint analysis based only on the
combination of the first three probes, namely: (i) SZE/X-ray
distances, (ii) the age of the oldest intermediate redshift
objects, and (iii) the measurements of the  Hubble Parameter
$H(z)$. Further, a complete joint analysis including the BAO
signature from the SDSS catalog it will be performed. For both
analysis, we stress that a specific flat $\Lambda$CDM cosmology
has not a priori been fixed.

All the observational expression adopted here have only two free
parameters ($\Omega_{M}, h$). In this way, we perform the
$\chi^{2}$ statistics over the $\Omega_{M}-h$ plane. Combining the
three tests discussed above the $\chi^{2}$ value reads:
\begin{eqnarray}
\chi^2(z|\mathbf{p})  = & \sum_i { ({\cal{D}}_A(z_i; \mathbf{p})-
{\cal{D}}_{A obs,i})^2 \over \sigma_{{\cal{D}}_{A obs,i}}^2 +
\sigma_{stat}^{2} + \sigma_{syst}^{2}} + \sum_j { (t(z_j;
\mathbf{p})-t_{inc}- t_{obs,j})^2 \over
\sigma_{t_{obs,j}}^2+\sigma_{t_{inc}}^2+\sigma_{syst}^{2}}
\nonumber
\\ & + \sum_k { (H(z_k; \mathbf{p})-H_{obs,k})^2 \over
\sigma_{H_{obs,k}}^2+\sigma_{syst}^{2}},
\end{eqnarray}
%\begin{equation}
%\chi^2(z|\mathbf{p}) = \sum_i { ({\cal{D}}_A(z_i; \mathbf{p})-
%{\cal{D}}_{A obs,i})^2 \over \sigma_{{\cal{D}}_{A obs,i}}^2 +
%\sigma_{stat}^{2} + \sigma_{syst}^{2}} + \sum_j { (t(z_j;
%\mathbf{p})-t_{inc}- t_{obs,j})^2 \over
%\sigma_{t_{obs,j}}^2+\sigma_{t_{inc}}^2+\sigma_{syst}} + \sum_k {
%(H(z_k; \mathbf{p})-H_{obs,k})^2 \over
%\sigma_{H_{obs,k}}^2+\sigma_{syst}},
%\end{equation}
where the quantities with subindex ``$obs$" are the observational
quantities, $\sigma_{{\cal{D}}_{Ao,i}}$ is the uncertainty in the
individual distance, $\sigma_{stat}$ is the contribution of the
statistical errors, $\sigma_{t_{inc}}$ is the incubation time
error, $\sigma_{syst}$ are the contribution of the systematic
errors for each sample added in quadrature and the complete set of
parameters is given by $\mathbf{p} \equiv (\Omega_{M}, h)$.

In Figure 2a, we show  dimensionless  $h$ parameter versus
the matter density parameter ($\Omega_M$). Our analysis combining
statistical and systematic errors from galaxy ages are in black
lines with $68.3\%$, $95.4\%$ and $99.7\%$ confidence levels
(c.l.). In the same way, the green lines represent the confidence
levels for the SZE/X-ray sample whereas the corresponding
constraints from Hubble parameter $H(z)$ sample are show in orange
lines. As should be expected,  the constraints of each sample on
$(\Omega_M, h)$ plane are too weak when individually considered.
However, due to the complementarity among them, our joint analysis
for these three samples predicts $H_0 = 73.4\pm 4.7$ km s$^{-1}$
Mpc$^{-1}$ and $\Omega_M=0.290^{+0078}_{-0.061}$
($1\sigma$ c.l.) for two free parameters with $\chi^2_{min}=
41.42$. The reduced values are $\chi^2_{red}= 0.702$ (including
systematics) and $\chi^2_{red} \cong 1$ (no systematics).

In Figure 2b, we display the contours on the space parameter
obtained through a joint analysis involving the combination of all
four probes. The dashed lines are cuts on $(\Omega_M, h)$ plane
from BAO signature. The red, green and blue contours are
constraints with $68.3\%$, $95.4\%$ and $99.7\%$ c.l.,
respectively. This complete analysis provides $H_0 =
74.1^{+3.3}_{-3.3}$ km.s$^{-1}$.Mpc$^{-1}$ ($4.5\%$ uncertainty)
whereas the density parameter is $\Omega_M=
0.278^{+0.034}_{-0.028}$ for two free parameters with a
$\chi^2_{min}= 41.53$ ($\chi^2_{red}= 0.704$).

\begin{table*}
\caption{Constraints on $h$ for different methods and epochs.}
\centerline{{\begin{tabular} {@{}cccc@{}}
Reference & Method & $h$ ($1\sigma$) & Epoch\\
\hline \hline Hinshaw {\it{et al.}} 2012
& WMAP-9 &$0.700\pm 0.022$ & $z\sim 1100$ \\
Ade {\it{et al.}} 2013
& PLANCK &$0.674\pm 0.014$ & $z\sim 1100$ \\
Riess {\it{et al.}} 2011& Cepheid+SNe+Maser &$0.738\pm 0.024$ & $z\simeq 0$ \\
Freedman {\it{et al.}} 2012& HST Key Project &$0.743\pm 0.026$ & $z\simeq 0$ \\
{\bf{This letter}}& {\bf{SZE/X-ray+Age+H(z)}}&$0.734\pm 0.031$ & $0.1 < z < 1.8$ \\
{\bf{This letter}}& {\bf{SZE/X-ray+Age+H(z)+BAO}}&$0.741\pm 0.022$ & $0.1 < z < 1.8$ \\
\hline
\end{tabular}} \label{ta2}}
\end{table*}

In Figure 2c, we show the likelihood function for the $h$
parameter in a flat $\Lambda$CDM universe. Both curves were
obtained by marginalizing on the matter density parameter. The shadow lines are cuts in the regions of $68.3\%$
and $95.4\%$ probability. For the solid black line the BAO
signature was not considered. The constraints for the black line
are $h= 0.734\pm 0.031$ ($0.064$) with $1\sigma$ ($2\sigma$),
respectively. For the red line the BAO signature has been
included. The constraints are  $h= 0.741\pm 0.022$ (corresponding
to $3\%$ error in $h$) and $0.045$ ($6.1\%$) with $1\sigma$ and
$2\sigma$, respectively. For all these analyses, the statistical
and systematic errors were added.

In Table 1, we compare the results derived here with the latest
determinations of $H_0$. Note that competitive
constraints were obtained using only probes at intermediate
redshifts, but the present results are nicely consistent with the
latest determinations based on the HST Key Project.

\section{Conclusions}\label{sec:Conclusions}

Several ongoing and future experiments (HST Key Project, S$H_0$ES,
PLANCK, GAIA, Spitzer/CHP, JWST, and others) are dedicated to
more accurately measuring the value of $H_0$, partially,  because an accuracy better
than $2\%$ will provide critical information on several cosmological
parameters.

We have demonstrated here that a joint analysis involving four
independent cosmological tests at intermediate redshifts ($z \sim
1$) provides an independent $3\%$ determination of $H_0$
(statistical and systematic errors combined). In the framework
of a flat $\Lambda$CDM model we have obtained  $h = 0.741 \pm
0.022$ (including statistical plus systematics), as shown
in table I, this value is not only consistent, but has the same
precision of a recent $H_0$ determination using nearby Cepheids
and Supernovae (Riess {\it{et al.}} 2011; Freedman {\it{et al.}}
2012). It is also suggested here that the present
determination of the Hubble constant based only on probes at
intermediate redshifts provides a competitive cross-check for any
determination of $H_0$.

The main advantage of the present treatment is that it does not
rely on extragalactic distance ladder being fully independent of
local calibrators. Naturally, its basic disadvantage rests on the
large and different systematic uncertainties appearing in each
probe when separately applied. However, the remarkable
complementarity among the four tests works in concert thereby
reducing greatly the possible degeneracy and uncertainties
appearing in the present determination of the Hubble constant.

Finally, concerning the intriguing tension between the $H_0$ determinations
from nearby objects and current CMB data, our analysis at
intermediate $z's$ is clearly favoring the local methods.
\begin{acknowledgments}
\noindent The authors acknowledge the referee for his/her suggestions that improved the 
manuscript.  JASL and JVC are partially supported by CNPq (Brazilian
Research Agency). 
\end{acknowledgments}

\end{document}